\documentstyle[sprocl]{article}

\input{psfig.sty}

\def\NCA{{\em Nuovo Cimento}}

\def\PRL{{\em Phys. Rev. Lett.}}
\def\PRD{{\em Phys. Rev.} D}
\def\ZPC{{\em Z. Phys.} C}

\def\be{\begin{equation}}
\def\ee{\end{equation}}
\def\bea{\begin{eqnarray}}
\def\eea{\end{eqnarray}}

\begin{document}

\begin{flushright}                                
UMN-D-98-3 \\ June 17, 1998 \\ \vspace{0.3in}
\end{flushright}

\title{PAULI--VILLARS REGULARIZATION IN A DISCRETE
LIGHT-CONE MODEL%
\footnote{To appear in the proceedings of               
the third workshop on Continuous Advances in QCD,
Minneapolis, Minnesota, April 16-19, 1998.}%
}

\author{J. R. Hiller}

\address{%
Department of Physics, 
University of Minnesota Duluth \\ 
Duluth, MN 55812, USA \\
E-mail: jhiller@d.umn.edu} 

\maketitle
\abstracts{%
Pauli--Villars regularization is successfully applied to 
nonperturbative calculations in a $(3+1)$-dimensional 
light-cone model.  Numerical results obtained with discretized
light-cone quantization compare favorably with the
analytic solution.}

\section{Introduction}

One of the challenges of using light-cone methods~\cite{Dirac,LCQreview}
to solve nonperturbative problems in field theories such as quantum
chromodynamics (QCD) is to develop a nonperturbative regularization and
renormalization scheme.  Here we discuss a new approach~\cite{PV,Wilson}
based on generalized Pauli--Villars regularization~\cite{PauliVillars}
and discretized light-cone quantization (DLCQ).\cite{PauliBrodsky}
The DLCQ Fock basis is expanded to include momentum states of
Pauli--Villars particles which provide the cancelations needed
to regulate ultraviolet divergences.  The determination of the
number of Pauli--Villars particles and of their coupling strengths
is done in perturbation theory, where loop integrals are rendered 
absolutely convergent.  

We test this approach on a simple model designed to have an
analytic solution for the lowest massive state.  The model is
related to an equal-time model of Greenberg and 
Schweber,\cite{GreenbergSchweber} which has also been translated
to the light cone by G{\l}azek and Perry.\cite{GlazekPerry}

\section{Light-Cone Coordinates}

We define light-cone coordinates and momentum components by~\cite{Dirac}
\begin{equation}  \label{eq:coordinates}
x^\pm=t\pm z\,,\;\;{\bf x}_\perp=(x,y)\,,
\end{equation}
and
\begin{equation}  \label{eq:momentum}
p^\pm=E\pm p_z\,,\;\;{\bf p}_\perp=(p_x,p_y)\,,\;\;
\underline{p}\equiv(p^+,{\bf p}_\perp)\,.
\end{equation}
The dot product is written
\begin{equation} 
p\cdot x=\frac{1}{2}(p^+x^-+p^-x^+)-{\bf p}_\perp\cdot{\bf x}_\perp\,.
\end{equation}
The time variable is taken to be $x^+$.  The conjugate variable
$p^-$ is the light-cone energy, and the associated operator ${\cal P}^-$
determines the time evolution of the system.
The mass-squared operator, frequently called the light-cone 
Hamiltonian,\cite{PauliBrodsky} is
\begin{equation}  \label{eq:HLC}
H_{\rm LC}={\cal P}^+{\cal P}^- - {\cal P}^2_\perp\,,
\end{equation}
where ${\cal P}^+$ and ${\bf \cal P}_\perp$ are momentum operators
conjugate to $x^-$ and ${\bf x}_\perp$.
In these coordinates, the fundamental eigenvalue problem is
\begin{equation}  \label{eq:EigenProb}
H_{\rm LC}\Psi=M^2\Psi\,,\;\; 
\underline{\cal P}\Psi=\underline{P}\Psi\,,
\end{equation}
where $M$ is the mass of the state $\Psi$.

The use of light-cone coordinates has several 
advantages.\cite{Dirac,LCQreview}
The generators of the Poincar\'{e} algebra have the
largest possible nondynamical subset;
in particular, boosts are kinematical.
The perturbative vacuum is the physical vacuum, because
$p^+=\sqrt{p^2+m^2}+p_z>0$, and there is then no need 
to compute the vacuum state.\footnote{For theories with complicated
vacuum structure in equal time formulations this simplicity
of the vacuum is offset by a more complicated operator
structure.\protect\cite{LCQreview}}  Fock-state expansions are well defined
with no disconnected vacuum pieces.

For field theories quantized in light-cone coordinates one
does need to be careful about the use of Pauli--Villars
regulators.  The number of Pauli--Villars particles used
in ordinary Feynman perturbation theory may not suffice.
One concrete example can be found in Yukawa 
theory.\cite{ChangYan}
To regulate the one-loop self-energy three
heavy bosons are required, rather than the usual one.  The reason
that one is not sufficient is that symmetric integration
techniques cannot be employed; the use of such techniques
in equal-time quantization amounts to a regularization
prescription which supplements the Pauli--Villars
regularization.

To see the need for three heavy bosons explicitly, consider
the self-energy integral
\begin{eqnarray}
I(\mu^2,M^2)&\equiv&-\frac{1}{\mu^2}\int\frac{dl^+d^2l_\perp}{l^+(q^+-l^+)^2} 
\nonumber \\
     & & \rule{0.5in}{0in}\times\frac{(q^+)^2{\bf l}_\perp^2+(2q^+-l^+)^2M^2}
           {M^2-D}\theta(\Lambda^2-D)\,,
\end{eqnarray}
where $\mu$ is the ordinary boson mass, $M$ is the fermion mass, and
$D$ is the invariant mass of the intermediate state
\begin{equation} D=\frac{\mu^2+{\bf l}_\perp^2}{l^+/q^+}
               +\frac{M^2+{\bf l}_\perp^2}{(q^+-l^+)/q^+}\,. 
\end{equation}
The integral has been made finite by an invariant-mass cutoff
$\Lambda^2$, and
can be done exactly; however, an expansion in powers of the fermion
mass is much more instructive.  We find
\begin{eqnarray}
I(\mu^2,M^2)&\simeq&
\frac{\pi}{\mu^2}\left[\left( \frac{\Lambda^2}{2} - \mu^2 \ln \Lambda^2
           + \mu^2 \ln \mu^2 - \frac{\mu^4}{2\Lambda^2}\right)\right. 
\nonumber \\
& +& M^2\left( 3 \ln\Lambda^2 - 3 \ln \mu^2 - \frac{9}{2}
                           + \frac{5\mu^2}{\Lambda^2}\right) 
\nonumber \\
& + & \left.M^4\left(\frac{2}{\mu^2}\ln (M^2/\mu^2)
                 +\frac{1}{3\mu^2}-\frac{1}{2\Lambda^2}\right)\right]\,.
\end{eqnarray}
The leading term violates the chiral symmetry of the
original theory; to remove this term from the infinite-cutoff
limit requires three Pauli-Villars bosons with different masses
$\mu_i$.  The subtracted integral
is
\begin{equation} 
I_{\rm sub}(\mu^2,M^2,\mu_i^2)=
   I(\mu^2,M^2)+\sum_{i=1}^3 C_i I(\mu_i^2,M^2)\,,\end{equation}
and the $C_i$ are chosen to satisfy
\begin{equation} 
1+\sum_{i=1}^3 C_i=0\,, \;\;
\mu^2+\sum_{i=1}^3 C_i\mu_i^2=0\,, \;\;
\sum_{i=1}^3 C_i\mu_i^2\ln(\mu_i^2/\mu^2)=0\,.
\end{equation}
Given this perturbative analysis of the regularization, the
conjecture is that fermion self-energies in
a nonperturbative calculation will be
regularized by the same Pauli--Villars bosons with the same
coupling strengths $C_i$.\footnote{To completely regularize Yukawa
theory will require a $\phi^4$ term and perhaps a
Pauli--Villars fermion.}

\section{A Soluble Model}

To test this idea, we have constructed a model remotely related
to Yukawa theory but which has the advantage of being analytically
soluble.  This type of model was investigated in equal-time
quantization by Greenberg and Schweber~\cite{GreenbergSchweber} 
and on the light cone by G{\l}azek and Perry.\cite{GlazekPerry}
It can be obtained from the Yukawa Hamiltonian~\cite{McCartorRobertson} 
by a number of severe modifications.
The momentum dependence in the fermion kinetic energy becomes
$(M_0^2+M'_0p^+)/P^+$, where $p^+$ and $P^+$ are the longitudinal
momenta of the fermion and system, respectively.\footnote{This term
has a structure similar to that of the self-induced inertia term
shown in Eq.~(C.2) of Ref.~\ref{McCartorRobertson}.}
Only the no-flip three-point vertex is kept and then only in a modified 
form where the longitudinal momentum dependence is simplified.
The structure of the chosen interactions results in states
populated by a fixed number of fermions and a cloud of bosons.
We study only the lowest state with one fermion.
One Pauli--Villars field is found to be sufficient in this case.
  
The resulting light-cone
Hamiltonian $H_{\rm LC}^{\rm eff}=P^+P_{\rm eff}^-$ is given by
\begin{eqnarray}
H_{\rm LC}^{\rm eff}&=&\int\frac{dp^+d^2p_\perp}{16\pi^3p^+}(M_0^2+M'_0p^+)
       \sum_\sigma b_{\underline{p}\sigma}^\dagger b_{\underline{p}\sigma} 
\nonumber \\
  & & +P^+\int\frac{dq^+d^2q_\perp}{16\pi^3q^+}
   \left[\frac{\mu^2+q_\perp^2}{q^+}a_{\underline{q}}^\dagger a_{\underline{q}}
     + \frac{\mu_1^2+q_\perp^2}{q^+}a_{1\underline{q}}^\dagger a_{1\underline{q}}
               \right]
 \\
   &  & +g\int\frac{dp_1^+d^2p_{\perp1}}{\sqrt{16\pi^3p_1^+}}
            \int\frac{dp_2^+d^2p_{\perp2}}{\sqrt{16\pi^3p_2^+}}
              \int\frac{dq^+d^2q_\perp}{16\pi^3q^+}
        \sum_\sigma b_{\underline{p}_1\sigma}^\dagger b_{\underline{p}_2\sigma}
\nonumber \\
    &  & \rule{0.25in}{0mm}\times \left[
      \left(\frac{p_1^+}{p_2^+}\right)^\gamma
    a_{\underline{q}}^\dagger\delta(\underline{p}_1-\underline{p}_2+\underline{q})
        +\left(\frac{p_2^+}{p_1^+}\right)^\gamma
   a_{\underline{q}}\delta(\underline{p}_1-\underline{p}_2-\underline{q}) \right.
\nonumber \\
    &  & \rule{0.25in}{0mm} \left.
       +i\left(\frac{p_1^+}{p_2^+}\right)^\gamma
   a_{1\underline{q}}^\dagger\delta(\underline{p}_1-\underline{p}_2+\underline{q})
      +i\left(\frac{p_2^+}{p_1^+}\right)^\gamma
a_{1\underline{q}}\delta(\underline{p}_1-\underline{p}_2-\underline{q}) \right]\,,
\nonumber 
\end{eqnarray}
where
\begin{equation}
\left[a_{\underline{q}},a_{\underline{q}'}^\dagger\right]
          =16\pi^3q^+\delta(\underline{q}-\underline{q}')\,,\;\;
\left\{b_{\underline{p}\sigma},b_{\underline{p}'\sigma'}^\dagger\right\}
     =16\pi^3p^+\delta(\underline{p}-\underline{p}')\delta_{\sigma\sigma'}\,.
\end{equation}
The state vector is written as a Fock-state expansion
\begin{eqnarray}
\Phi_\sigma&=&\sqrt{16\pi^3P^+}\sum_{n,n_1}\int\frac{dp^+d^2p_\perp}{\sqrt{16\pi^3p^+}}
   \prod_{i=1}^n\int\frac{dq_i^+d^2q_{\perp i}}{\sqrt{16\pi^3q_i^+}}
   \prod_{j=1}^{n_1}\int\frac{dr_j^+d^2r_{\perp j}}{\sqrt{16\pi^3r_j^+}} 
\nonumber \\
   &  & \times \delta(\underline{P}-\underline{p}
                     -\sum_i^n\underline{q}_i-\sum_j^{n_1}\underline{r}_j)
       \phi^{(n,n_1)}(\underline{q}_i,\underline{r}_j;\underline{p}) 
\nonumber \\
   &  &  \rule{0.5in}{0mm} \times 
         \frac{1}{\sqrt{n!n_1!}}b_{\underline{p}\sigma}^\dagger
          \prod_i^n a_{\underline{q}_i}^\dagger 
             \prod_j^{n_1} a_{1\underline{r}_j}^\dagger |0\rangle \,,
\end{eqnarray}
with $n$ the number of ordinary bosons and $n_1$ the number of 
Pauli--Villars bosons.  The normalization of the state is
chosen to be
\begin{equation} 
\Phi_\sigma^{\prime\dagger}\cdot\Phi_\sigma
=16\pi^3P^+\delta(\underline{P}'-\underline{P})\,. 
\end{equation}
For the boson amplitudes this implies
\begin{eqnarray}
1&=&\sum_{n,n_1}\prod_i^n\int\,dq_i^+d^2q_{\perp i}
                     \prod_j^{n_1}\int\,dr_j^+d^2r_{\perp j} 
\nonumber \\
    & & \times \left|\phi^{(n,n_1)}(\underline{q}_i,\underline{r}_j;
           \underline{P}-\sum_i\underline{q}_i-\sum_j\underline{r}_j)\right|^2\,.
\end{eqnarray} 
             
A solution to the mass eigenvalue problem
\begin{equation}
H_{\rm LC}^{\rm eff}\Phi_\sigma=M^2\Phi_\sigma 
\end{equation}
must satisfy the following coupled set of integral equations:
\begin{eqnarray}  \label{eq:CoupledEqns}
\lefteqn{\left[M^2-M_0^2-M'_0p^+
  -\sum_i\frac{\mu^2+q_{\perp i}^2}{y_i}
                  -\sum_j\frac{\mu_1^2+r_{\perp j}^2}{z_j}\right]
                    \phi^{(n,n_1)}(\underline{q}_i,
                                       \underline{r}_j,\underline{p}) } 
\nonumber \\
& & =g\left\{\sqrt{n+1}\int\frac{dq^+d^2q_\perp}{\sqrt{16\pi^3q^+}}
              \left(\frac{p^+-q^+}{p^+}\right)^\gamma
              \phi^{(n+1,n_1)}(\underline{q}_i,\underline{q},
                              \underline{r}_j,\underline{p}-\underline{q})\right.
\nonumber \\
& &\rule{0.5in}{0mm} +\frac{1}{\sqrt{n}}\sum_i\frac{1}{\sqrt{16\pi^3q_i^+}}
              \left(\frac{p^+}{p^++q_i^+}\right)^\gamma
 \\
& &\rule{0.75in}{0mm} \times
              \phi^{(n-1,n_1)}(\underline{q}_1,\ldots,\underline{q}_{i-1},
                                 \underline{q}_{i+1},\ldots,\underline{q}_n,
                                  \underline{r}_j,\underline{p}+\underline{q}_i)
\nonumber \\
& &\rule{0.5in}{0mm}+i\sqrt{n_1+1}\int\frac{dr^+d^2r_\perp}{\sqrt{16\pi^3r^+}}
              \left(\frac{p^+-r^+}{r^+}\right)^\gamma
              \phi^{(n,n_1+1)}(\underline{q}_i,\underline{r}_j,
                                     \underline{r},\underline{p}-\underline{r})
\nonumber \\
& &\rule{0.5in}{0mm}+\frac{i}{\sqrt{n_1}}\sum_j\frac{1}{\sqrt{16\pi^3r_j^+}}
              \left(\frac{p^+}{p^++r_j^+}\right)^\gamma
\nonumber \\
& &\rule{0.75in}{0mm} \times\left.
              \phi^{(n,n_1-1)}(\underline{q}_i,\underline{r}_1,\ldots,\underline{r}_{j-1},
                                 \underline{r}_{j+1},\ldots,\underline{r}_{n_1},
                                     \underline{p}+\underline{r}_j) \right\}\,.
\nonumber
\end{eqnarray}
The solution is
\begin{eqnarray}
\phi^{(n,n_1)}&=&\sqrt{Z}\frac{(-g)^n(-ig)^{n_1}}{\sqrt{n!n_1!}}
                 \left(\frac{p^+}{P^+}\right)^\gamma
                  \prod_i\frac{y_i}{\sqrt{16\pi^3q_i^+}(\mu^2+q_{\perp i}^2)} 
\nonumber \\
        & & \rule{0.5in}{0mm} \times  
                \prod_j\frac{z_j}{\sqrt{16\pi^3r_j^+}(\mu_1^2+r_{\perp j}^2)}\,.
\end{eqnarray}
provided that $M_0=M$ and
\begin{equation}
M'_0=\frac{g^2/P^+}{16\pi^2}\frac{\ln\mu_1/\mu}{\gamma+1/2}\,.
\end{equation}
Given the structure of the individual amplitudes, a natural
choice for the value of the interaction parameter $\gamma$
is seen to be 1/2, because each amplitude is then proportional
to the square root of the product of all longitudinal momenta.
With this value of 1/2, the 
wave function normalization can be reduced to a rapidly converging
infinite sum
\begin{equation} 
\frac{1}{Z}=\sum_{n,n_1}^\infty \frac{1}{(2n+2n_1+1)!n!n_1!}
              \frac{(g/\mu)^{2n} (g/\mu_1)^{2n_1}}{(16\pi^2)^{n+n_1}}\,, 
\end{equation}

To fix the coupling $g$ we set the value of the
expectation value $\langle :\!\!\phi^2(0)\!\!:\rangle
\equiv\Phi_\sigma^\dagger\!:\!\!\phi^2(0)\!\!:\!\Phi_\sigma$.
For the analytic solution (with $\gamma=1/2$) it reduces to
\begin{equation}
\langle :\!\!\phi^2(0)\!\!:\rangle=
      \sum_{n,n_1}^\infty \frac{2Zn}{(2n+2n_1)!n!n_1!}
      \frac{(g/\mu)^{2n} (g/\mu_1)^{2n_1}}{(16\pi^2)^{n+n_1}}  \,.
\end{equation}
From a numerical solution the quantity
can be computed fairly efficiently in a sum
similar to the normalization sum
\begin{eqnarray}
\langle :\!\!\phi^2(0)\!\!:\rangle
        &=&\sum_{n=1,n_1=0}\prod_i^n\int\,dq_i^+d^2q_{\perp i} 
                \prod_j^{n_1}\int\,dr_j^+d^2r_{\perp j}
  \\
    & & \rule{0.25in}{0mm} \times \left(\sum_{k=1}^n \frac{2}{q_k^+/P^+}\right)
              \left|\phi^{(n,n_1)}(\underline{q}_i,\underline{r}_j;
       \underline{P}-\sum_i\underline{q}_i-\sum_j\underline{r}_j)\right|^2\,,
\nonumber
\end{eqnarray}
with the integrals computed from discrete approximations.

As a prediction of the model, we compute a distribution function
for the physical bosons
\begin{eqnarray}
f_B(y)&\equiv&\sum_{n,n_1}\prod_i^n\int\,dq_i^+d^2q_{\perp i}
                     \prod_j^{n_1}\int\,dr_j^+d^2r_{\perp j} 
     \sum_{i=1}^n\delta(y-q_i^+/P^+)  
\nonumber \\
&&\rule{0.5in}{0mm}\times
    \left|\phi^{(n,n_1)}(\underline{q}_i,\underline{r}_j;
           \underline{P}-\sum_i\underline{q}_i-\sum_j\underline{r}_j)\right|^2\,.
\end{eqnarray}
and the average multiplicity
\begin{equation}
\langle n_B\rangle=\int_0^1f_B(y)dy\,.
\end{equation}
For the analytic solution we obtain
\begin{equation}
f_B(y)=\sum_{n,n_1}^\infty \frac{Zny(1-y)^{(2n+2n_1-1)}}{(2n+2n_1-1)!n!n_1!}
                         \frac{(g/\mu)^{2n} (g/\mu_1)^{2n_1}}{(16\pi^2)^{n+n_1}}
\end{equation}
and
\begin{equation}
\langle n_B\rangle=
    \sum_{n,n_1}^\infty \frac{Zn}{(2n+2n_1+1)!n!n_1!}
              \frac{(g/\mu)^{2n} (g/\mu_1)^{2n_1}}{(16\pi^2)^{n+n_1}}\,.
\end{equation}

Another prediction is the slope of the form factor for the dressed fermion.
An expression can be constructed~\cite{PV} from the formalism developed
by Brodsky and Drell.\cite{BrodskyDrell}  For the analytic solution
of the model we obtain
\begin{equation}
F'(0)=-\sum_{n,n_1} \frac{Z(n/\mu^2+n_1/\mu_1^2)}{(2n+2n_1+3)!n!n_1!}
          \frac{(g/\mu)^{2n} (g/\mu_1)^{2n_1}}{(16\pi^2)^{n+n_1}}\,.
\end{equation}
Numerically, one can estimate $F'(0)$ from
\begin{eqnarray} \label{eq:BetterFprime}
\lefteqn{\tilde{F}'(0)=-\sum_{n,n_1}\prod_i^n\int\,dq_i^+d^2q_{\perp i}
                   \prod_j^{n_1}\int\,dr_j^+d^2r_{\perp j}}\hspace{0.5in} \\
     & \times & \left[\sum_i \left|\frac{y_i}{2}\nabla_{\perp i}
             \phi^{(n,n_1)}(\underline{q}_i,\underline{r}_j;
           \underline{P}-\sum_i\underline{q}_i-\sum_j\underline{r}_j)
                                        \right|^2 \right.
    \nonumber \\
    & & \left. +\sum_j \left|\frac{z_j}{2}\nabla_{\perp j}
                   \phi^{(n,n_1)}(\underline{q}_i,\underline{r}_j;
           \underline{P}-\sum_i\underline{q}_i
              -\sum_j\underline{r}_j)\right|^2 \right]\,,
    \nonumber
\end{eqnarray}
which differs from $F'(0)$ by surface terms which vanish as
$\Lambda\rightarrow\infty$.

\section{Numerical methods and results}

We solve the model numerically by using
DLCQ.\cite{PauliBrodsky}   We impose
periodic boundary conditions for bosons and antiperiodic conditions 
for fermions in a light-cone box $-L<x^-<L$, $-L_\perp<x,y<L_\perp$.
This introduces a discrete grid:
\begin{equation} p^+\rightarrow\frac{\pi}{L}n\,,\;\;
{\bf p}_\perp\rightarrow(\frac{\pi}{L_\perp}n_x,\frac{\pi}{L_\perp}n_y)\,.
\end{equation}
Integrals are replaced by discrete sums
\begin{equation} 
\int dp^+ \int d^2p_\perp f(p^+,{\bf p}_\perp)\simeq
   \frac{2\pi}{L}\left(\frac{\pi}{L_\perp}\right)^2
   \sum_n\sum_{n_x,n_y=-N_\perp}^{N_\perp}
   f(n\pi/L,{\bf n}_\perp\pi/L_\perp)\,. 
\end{equation}
The limit $L\rightarrow\infty$ can be exchanged for a limit
in terms of the integer {\em resolution}~\cite{PauliBrodsky}
\begin{equation}  
K\equiv\frac{L}{\pi}P^+\,.  
\end{equation}
Longitudinal momentum fractions are given by
$x=p^+/P^+\rightarrow n/K$, with n odd for fermions and 
even for bosons.  The light-cone Hamiltonian
$H_{\rm LC}$ is independent of $L$, but does depend on $K$.  

Because all $n$ are positive, DLCQ
automatically limits the number of particles to no more than $\sim K/2$.
The integers $n_x$ and $n_y$ range between limits associated with
some maximum integer $N_\perp$ fixed by the invariant-mass cutoff
\begin{equation} 
\frac{m_i^2+p_{\perp i}^2}{x_i}\leq\Lambda^2
\end{equation}
imposed for each constituent.
The eigenvalue problem (\ref{eq:CoupledEqns}) is then converted to a 
finite matrix problem.  Typical basis sizes are given in
Table~\ref{tab:basis50}.  The Hamiltonian matrix is quite sparse.
The lowest eigenvalue of the matrix is extracted with use of
the Lanczos algorithm~\cite{Lanczos}
for complex symmetric matrices.\footnote{The 
imaginary couplings of the Pauli--Villars particles make the 
Hamiltonian matrix complex 
symmetric.}

\begin{figure}[p]
\centerline{\psfig{figure=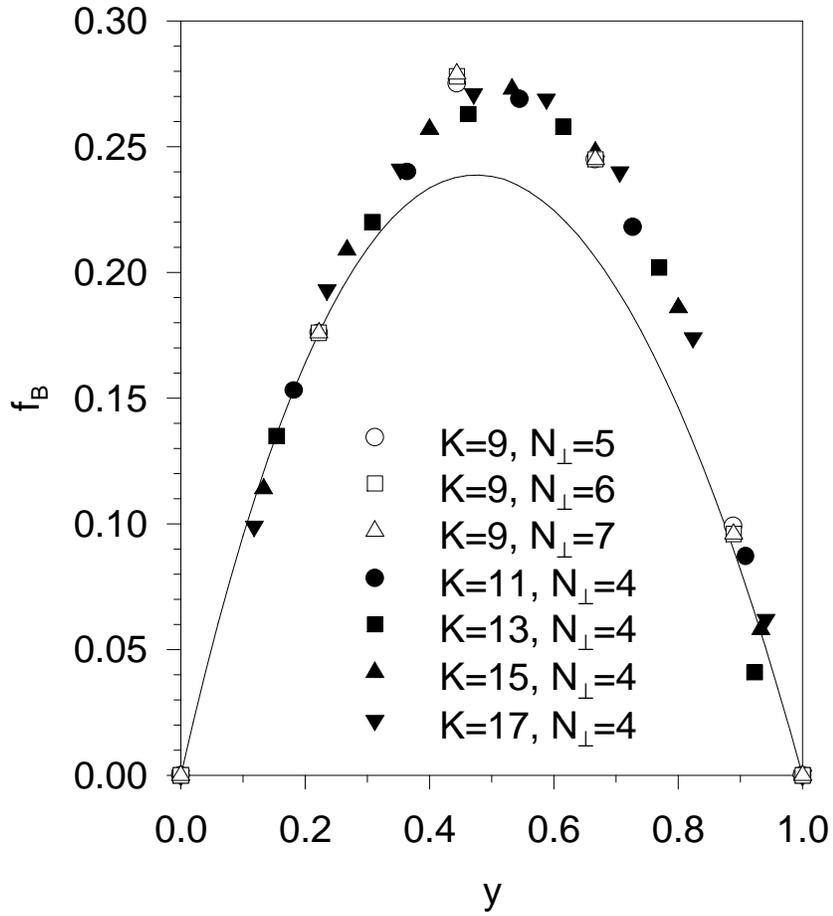,width=4.7in}}
\caption{%
The boson distribution function $f_B$ at various numerical resolutions,
with \mbox{$\langle:\!\!\phi^2(0)\!\!:\rangle=1$}
and $\Lambda^2=50\mu^2$.  The solid line is the analytic result at infinite
$\Lambda^2$.
\label{fig:fB-resolution}}
\end{figure}

\begin{figure}[p]
\centerline{\psfig{figure=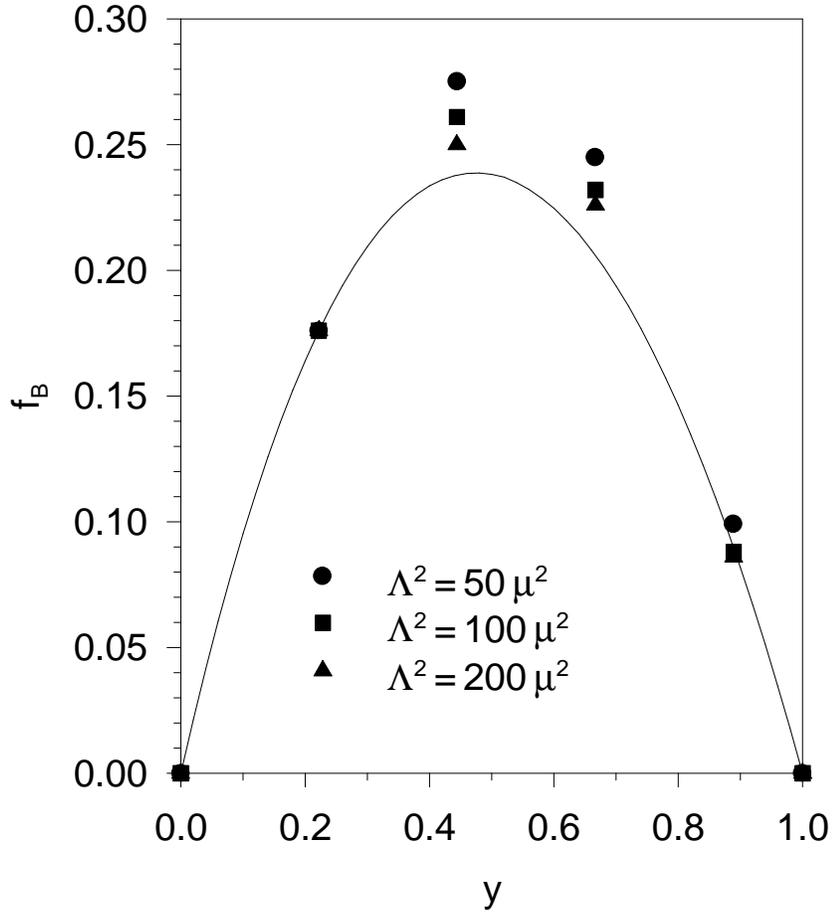,width=4.7in}}
\caption{%
The boson distribution function $f_B$ for different cutoff values,
with $\langle:\!\!\phi^2(0)\!\!:\rangle=1$ and numerical resolution
set at $K=9$ and $N_\perp=5$.  The solid line is the analytic result at
$\Lambda^2=\infty$.
\label{fig:fB-Lambda2}}
\end{figure}

Some of the results obtained are listed in Table~\ref{tab:phi1}
and displayed in Figs.~\ref{fig:fB-resolution} and \ref{fig:fB-Lambda2}.
Figure~\ref{fig:fB-resolution} shows that the results are quite
insensitive to numerical resolution, while Fig.~\ref{fig:fB-Lambda2}
illustrates how the analytic solution is approached as the cutoff is
increased.
These results demonstrate that the DLCQ approximation yields a 
good representation of the solution to the model.

\section{Summary}

We have developed and used a simple soluble model
to test the feasibility of Pauli--Villars regularization in DLCQ.
The number of Pauli--Villars Fock states required is not
prohibitive and good results are obtained.  From this
success we can work toward approximation of Yukawa theory
itself by adding fermion dynamics and gradually reinstating
the full complexity of the interactions.  We can also consider
application to other theories, perhaps including QCD.

\newpage
\begin{table}[t]
\caption{\label{tab:basis50}
Basis sizes for DLCQ calculations in the soluble model
with parameters $M^2=\mu^2$, $\mu_1^2=10\mu^2$, and 
$\Lambda^2=50\mu^2$.  The numbers of physical states
are in parentheses.}
\begin{tabular}{|c|rrrrrr|}
\hline 
{} & \multicolumn{6}{c|}{$K$} \\
 \cline{2-7}
$N_\perp$& 
      7  &    9   &    11   &    13   &     15  &     17  \\
\hline
1&    18 &     38 &      36 &      65 &     110 &    185 \\
 &   (7) &   (12) &    (19) &     (30)&     (45)&    (67)\\
2&   218 &    265 &     590 &    1120 &     822 &   1410 \\
 &  (127)&   (119)&    (343)&    (754)&    (453)& (626)  \\
3&   958 &   1408 &    4460 &   17031 &   22486 &  21635 \\
 &  (367)&   (736)&   (2671)&   (9230)&  (13213)& (13531)\\
4&  3714 &   9259 &   49394 &   50966 &  110254 & 328966 \\
 & (1399)&  (5913)&  (32363)&  (32124)&  (55319)&(172247)\\
5& 13702 &  54100 &   95176 &  386140 & 1553576 &  {}    \\
 & (5699)& (28065)&  (66371)& (232400)&(1038070)&  {}    \\
6& 35666 & 126748 &  536758 & 2907158 &         &  {}    \\
 &(12991)& (69245)& (391511)&(2107688)&         &  {}    \\
7& 79794 & 519325 & 1317392 &         &         &  {}    \\
 &(32891)&(276299)&(1008539)&         &         &  {}    \\
8&172118 &1165832 &         &         &         &  {}    \\
 &(61947)&(687394)&         &         &         &  {}    \\
\hline
\end{tabular}
\end{table}
\begin{table} 
\caption{\label{tab:phi1}
Numerical parameter values and results from solving 
the model eigenvalue problem.  The physical parameter
values were $M^2=\mu^2$ for the fermion mass, $\mu_1^2=10\mu^2$
for the Pauli--Villars mass, and $\langle:\!\!\phi^2(0)\!\!:\rangle=1$
to fix the coupling $g$.}
\begin{center}
\begin{tabular}{|cccccccc|}
\hline
$(\Lambda/\mu)^2$ & $K$ & $N_\perp$ & $\mu L_\perp/\pi$ & 
  $(M_0/\mu)^2$ & $g/\mu$ & $\langle n_B\rangle$ & $100\mu^2\tilde{F}'(0)$ \\
\hline
 50 & 11 & 4 & 0.8165 & 0.8547 & 13.293 & 0.177 & -0.751 \\
 50 & 13 & 4 & 0.8165 & 0.8518 & 13.230 & 0.172 & -1.015 \\
 50 & 15 & 4 & 0.8165 & 0.8408 & 13.556 & 0.178 & -0.715 \\
 50 & 17 & 4 & 0.8165 & 0.8289 & 13.392 & 0.180 & -0.565 \\
 {}&{}&{}&{}&{}&{}&{}&{}\\
 50 &  9 & 5 & 1.2062 & 0.8601 & 14.023 & 0.179 & -0.547 \\
 50 &  9 & 6 & 1.2247 & 0.8377 & 14.323 & 0.179 & -0.582 \\
 50 &  9 & 7 & 1.4289 & 0.8302 & 14.386 & 0.179 & -0.658 \\
 {}&{}&{}&{}&{}&{}&{}&{}\\
 50 &  9 & 5 & 1.2062 & 0.8601 & 14.023 & 0.179 & -0.547 \\
100 &  9 & 5 & 0.7143 & 1.0520 & 12.565 & 0.174 & -0.239 \\
200 &  9 & 5 & 0.5025 & 1.1980 & 10.191 & 0.172 & -0.139 \\
 {}&{}&{}&{}&{}&{}&{}&{}\\
$\infty$  &  \multicolumn{3}{c}{analytic} & 1.0000 & 13.148 & 0.160 & -0.786 \\
\hline
\end{tabular}
\end{center}
\end{table}

\clearpage
\section*{Acknowledgments}

The work reported here was done in collaboration 
with S.J. Brodsky and G. McCartor and was supported
in part by grants of computing time by the Minnesota
Supercomputer Institute.

\end{document}